\documentclass{article}

     \PassOptionsToPackage{numbers, compress, sort}{natbib}

     \usepackage[final]{neurips_2020}

\usepackage[utf8]{inputenc} %
\usepackage[T1]{fontenc}    %
\usepackage{url}            %
\usepackage{booktabs}       %
\usepackage{amsfonts}       %
\usepackage{nicefrac}       %
\usepackage{microtype}      %
\usepackage{geometry}
\usepackage[compact]{titlesec}
\usepackage{wrapfig}

\usepackage[pdftex]{graphicx}
\usepackage{url}
\usepackage{indentfirst}
\usepackage{listings}
\usepackage[fleqn]{amsmath}
\usepackage{amssymb}
\usepackage{graphicx}
\usepackage{subfig}
\usepackage[notes=true,later=false,done=false]{dtrt}
\usepackage{mathtools}

\usepackage{algorithm}
\usepackage[noend]{algpseudocode}
\MakeRobust{\Call}

\usepackage{tikz}
\usetikzlibrary{backgrounds}
\usepackage{pgfplots}
\usepackage{csvsimple}

\newcommand{\norm}[1]{\left\lVert#1\right\rVert}
\newcommand{\brac}[1]{\left(#1\right)}
\newcommand{\curbrac}[1]{\left\{#1\right\}}
\newcommand{\sqbrac}[1]{\left[#1\right]}

\newcommand{\size}[1]{\left|#1\right|}

\newcommand{\ideal}{\mathcal{I}}

\DeclareMathOperator{\argmax}{argmax}

\newcommand{\paragraphbe}[1]{\vspace{0.75ex}\noindent{\bf \em #1} }

\title{De-Anonymizing Text by \\ Fingerprinting Language Generation}

\author{%
  Zhen Sun\\
  Cornell University\\
  \texttt{zs352@cornell.edu} \\
   \And
   Roei Schuster \\
  Cornell Tech, Tel Aviv University \\
  \texttt{rs864@cornell.edu} \\
   \And
  Vitaly Shmatikov \\
  Cornell Tech \\
   \texttt{shmat@cs.cornell.edu} \\
}

\begin{document}

\maketitle

\begin{abstract}

Components of machine learning systems are not (yet) perceived as
security hotspots.  Secure coding practices, such as ensuring that no
execution paths depend on confidential inputs, have not yet been adopted
by ML developers.  We initiate the study of code security of ML systems
by investigating how nucleus sampling\textemdash a popular approach for
generating text, used for applications such as auto-completion\textemdash
unwittingly leaks texts typed by users.  Our main result is that the
series of nucleus sizes for many natural English word sequences is a
unique \emph{fingerprint}.  We then show how an attacker can infer typed
text by measuring these fingerprints via a suitable side channel (e.g.,
cache access times), explain how this attack could help de-anonymize
anonymous texts, and discuss defenses.

\end{abstract}

\section{Introduction}

Machine learning (ML) models are composed from building blocks such
as layer types, loss functions, sampling methods, etc.  Each building
block typically has a few popular library implementations, which are
incorporated into many models\textemdash including models whose inputs
are sensitive (e.g., private images or typed text).  Therefore, ML models
are ``security hotspots'' and their implementations must follow secure
coding practices.  This includes protecting the inputs from \emph{side
channels}, i.e., low-level physical or microarchitectural side effects
of the computation that are externally observable and leak information
about its internal state to concurrent, adversarial processes.

We use \emph{nucleus sampling}~\cite{holtzman2019curious}, a leading
approach for efficiently generating high-fidelity text, as a case study of
side-channel vulnerabilities in ML models.  Given the output probabilities
of a language model such as GPT-2~\cite{radford2019language}, nucleus
sampling draws candidates from a variable-sized ``nucleus'' of the
most probable words.  It is the basis of applications such as text
auto-completion~\cite{Wolf2019HuggingFacesTS, writewithtransformer}.

First, we demonstrate that \textbf{the series of nucleus sizes produced
when generating an English-language word sequence is a fingerprint} by
showing that the nucleus size series of any sequence satisfying a simple
criterion is far from any other sequence, unless their textual contents
substantially overlap.  We then derive a lower bound on the Euclidean
distance between fingerprints that depends only on the sequence length
but not on the size or domain of the corpus.

Second, we show that implementations of nucleus sampling, such as the
popular Hugging Face Transformers package, contain a dangerous information
leak.  An attacker who runs a concurrent, sandboxed application process
on the user's device can infer the nucleus size by indirectly measure
the number of iterations of a certain loop, and thus fingerprint the
input text.  We use Flush+Reload~\cite{yarom2014flush+} for our proof
of concept, but the general approach works with any suitable side
channel~\cite{gullasch2011cache, osvik2006cache, lipp2016armageddon}.

We design a fingerprint matching algorithm and show that (1) it tolerates
noise in side-channel measurements, and (2) does not produce false
positives.  Therefore, an attacker can accurately identify the typed
sequence out of many billions of possible candidates in an ``open-world''
setting, without assuming \textit{a priori} that the user's input belongs
to a known small dataset.  This technique can help de-anonymize text by
asynchronously matching fingerprints collected from the user's device
to anonymous blog entries, forum posts, emails, etc.  For example, we
show that many of the anonymous users' posts on the infamous Silk Road
forum have unique fingerprints.

We conclude by explaining how to mitigate the information leak and
discuss the importance of removing insecure coding patterns such as
input-dependent loops from ML building blocks.

\paragraphbe{Ethics and responsible disclosure.}
The purpose of this study is to improve the security of popular ML systems
and help protect the privacy of their users.  We disclosed our findings
and our proposed mitigation code by email to members of the Hugging Face
engineering team responsible for the implementation of nucleus sampling
(identified via a contact at Hugging Face and GitHub's commit log)
and a message to Hugging Face's public Facebook contact point.

We use Silk Road posts as a case study only because they represent
informal textual communications whose authors likely wish to maintain
their anonymity.  \textbf{Silk Road posts include offensive and harmful
content.  We use this dataset solely for our proof-of-concept experiments.
It does not reflect our views in any way.}

\section{Background}
\label{sec:background}
\label{sec:tgeneration}

\subsection{Text generation via language model sampling}
\label{sec:sampling}

Let $\mathbb{D}$ be a dictionary, $\mathbb{S}=\cup_{i\in
\mathbb{N}}\mathbb{D}^i$ a set of possible \textit{texts} (sequences
of dictionary words), and $X\in \mathbb{S}$.  A \textit{language model}
$\mathcal{M}:\mathbb{S}\rightarrow \mathbb{R}^{\size{\mathbb{D}}}$ maps
a ``prefix'' $\brac{x_1, \ldots, x_n}\in \mathbb{S}$ to a probability
distribution $\brac{p_1, \ldots p_{\size{\mathbb{D}}}}$ of the next word.
\textbf{Text auto-completion} is a popular application of language
generation.  As the user is typing some text $X\in \mathbb{S}$, a language
model is sampled at each time step $t\in\curbrac{1,..,\size{X}}$, to
generate a ``probable'' suffix for $X\sqbrac{:t}$ (the prefix of $X$
up to index $t$).

\textit{Pure} sampling draws the next word $y$ according to the
probabilities given by $\mathcal{M}\brac{x_1, \ldots x_n}$, then invokes
$\mathcal{M}$ on $\brac{x_1, \ldots, x_n, y}$, and so on.  Typically,
sampling stops when a special end-of-sequence or end-of-sentence token is
sampled, or when the probability of the entire sampled sequence (estimated
by multiplying the model's output probabilities for the sampled words)
drops below a certain threshold.  Other approaches include \textit{greedy}
sampling, which simply sets $x_{n+1}\gets \argmax{\mathcal{M}\brac{x_1,
\ldots, x_n}}$, and \textit{top-k} sampling, which selects words
corresponding to the top $k$ highest values in $\mathcal{M}\brac{x_1,
\ldots x_n}$ and applies pure sampling to them according to their
probabilities (normalized to sum up to 1).  Different sampling methods
generate text with different properties~\cite{holtzman2019curious,
welleck2020consistency}.  Pure sampling produces poor-quality text (often,
incomprehensible gibberish) as perceived by humans, while greedy sampling
results in a lack of language diversity, often with highly unnatural
repetition.

\textbf{Nucleus sampling}~\cite{holtzman2019curious} is similar to
top-k sampling but instead of choosing candidates based on ranks, it
chooses the maximal set (``nucleus'') of top-ranked words such that
the sum of their probabilities is $\leq q$.  It produces high-quality,
high-diversity text~\cite{holtzman2019curious} and performs well on
metrics, including the Human Unified with Statistical Evaluation (HUSE)
score~\cite{hashimoto2019unifying}.

\subsection{Microarchitectural side channels}
\label{sec:microattacks}

Process isolation in modern systems is a leaky abstraction.  If a
user's process and an attacker's concurrent process share physical
hardware resources, the attacker can infer information about the user's
activity by analyzing contention patterns on the cache (see below), cache
directories~\cite{yan2019attack}, GPU~\cite{naghibijouybari2018rendered},
translation lookaside buffer~\cite{gras2018translation}, and many
other resources.  These attacks, known as \textit{microarchitectural
side channels}, can be exploited by any untrusted, low-privilege
process.  Side-channel attacks have been demonstrated on many
PC and mobile~\cite{lipp2016armageddon} platforms, and even from
Javascript or WebAssembly code within the highly restricted browser
sandbox~\cite{oren2015spy, genkin2018drive}.

\label{sec:whocanrun}

Several programming patterns are especially vulnerable to side-channel
attacks.  Loop arguments are a textbook example~\cite{kocher1996timing}:
loops take longer to execute than non-iterative code, their execution
time can be inferred using coarse timers, and their side effects on
microarchitectural resources are repeated many times, amplifying the
signal.  Loops whose iterations depend on some secret can leak this secret
through many microarchitectural~\cite{liu2015last, gras2018translation,
yarom2014flush+} and physical~\cite{genkin2014rsa, genkin2016ecdh}
side channels.

\paragraphbe{Cache side channels.}
\label{sec:cacheattacks1}
Cache memory is shared even among isolated processes.  When the contents
of physical memory addresses are loaded or evicted by any process,
it affects how fast that memory can be accessed by other processes.
Therefore, memory access times measured by one process can reveal
which memory addresses are accessed by another process.  Cache attacks
have been used to extract cryptographic keys~\cite{zhang2012cross,
liu2015last, yarom2014flush+, osvik2006cache, percival2005cache,
cohney2020pseudorandom}, steal login tokens~\cite{ronen20199},
defeat OS security mechanisms~\cite{hund2013practical}, sniff user
inputs~\cite{lipp2016armageddon}, and more~\cite{zhang2012cross,
ristenpart2009hey, zhang2014cross}.

\emph{Flush+Reload}~\cite{yarom2014flush+, gullasch2011cache} is a popular
type of cache attacks.  When a victim process and a concurrent attacker
process load the same shared library or a file, a single set of physical
memory addresses containing the file's content is mapped into both
processes' virtual address space.  In this situation, the attacker can (1)
cause the eviction of a specific memory address (``flush''), (2) wait,
and (3) reload this memory address.  Short reload time reveals that the
victim has accessed this address after the eviction but before the load.
If memory addresses monitored by the attacker do not contain shared
memory, other cache attacks such as Prime+Probe~\cite{osvik2006cache,
percival2005cache} may be used instead.

\section{Fingerprinting auto-completed text sequences}
\label{sec:fingerprint}

Consider a text auto-completion assistant that uses nucleus sampling
(see Section~\ref{sec:sampling}).  At each step $t$, the user has typed
$X\sqbrac{:t}$.  The assistant uses $\mathcal{M}\brac{X\sqbrac{:t}}$
to generate nucleus $q$ and samples from it to auto-complete the user's
text. The user may accept the completion or manually type in the next word.
 For texts that are typed over multiple ``sessions,'' e.g., in
separate forum posts, we assume the assistant stops sampling at the
end of every text and resets.  Let $\ideal_{\mathcal{M},q}\brac{X}\in
\mathbb{R}^{\size{X}}$ be the resulting \textit{nucleus size series}
(NSS).  Figure \ref{fig:fingerprint} shows an example.

\subsection{Fingerprints of text sequences}

Let $X,Y \in \mathbb{S}$ be text sequences s.t.\ $\size{X}=\size{Y}$.
We say that $X$ and $Y$ are \textit{similar} if they have
identical subsequences of length $N$ starting at the same index,
i.e., $\exists i\in \mathbb{Z}$, $0 \leq i < \size{X} - N$ s.t.\
$X\sqbrac{i:i+N}=Y\sqbrac{i:i+N}$.  We set $N=50$, which is a very
rigorous criterion for similarity: if two sequences have a common 50-word
subsequence in exactly the same position, they are likely identical or
have some identical source (and are thus \emph{semantically} close to
each other).

Let $\pi$ be a procedure that receives as input $X\in\mathbb{S}$ and
returns a vector in $\mathbb{R}^{\size{X}}$.  We say that $\pi\brac{X}$
is a \textbf{fingerprint} if there exists a monotonically increasing
\emph{uniqueness radius} $U:\mathbb{N}\rightarrow \mathbb{R}$
such that for any $Y\in \mathbb{S}$ which is not similar to $X$ and
$\size{Y}=\size{X}$, $\norm{\pi\brac{X}-\pi\brac{Y}}>U\brac{\size{X}}$
where $\norm{.}$ is the Euclidean norm.  In other words, there exists
a ``ball'' around the fingerprint of any sequence $X$ such that no
other sequence has its fingerprint within that ball (unless it is
similar to $X$).  This defines an \emph{open-world} fingerprint, i.e.,
the uniqueness of a sequence's fingerprint holds with respect to all
natural-language sequences and not just a specific dataset.

\subsection{Nucleus size series is a fingerprint}
\label{sec:emp}
\label{sec:defs}
\vspace{-0.07em}
We conjecture that $\pi\brac{X} = \ideal_{\mathcal{M},q}\brac{X}$
of any English sequence is a fingerprint, as long as
$\ideal_{\mathcal{M},q}\brac{X}$ is sufficiently ``variable.''
We define variability of $\ideal_{\mathcal{M},q}\brac{X} =
(n_1, \ldots, n_{\size{X}})$ as $\sqrt{\frac{1}{\size{X}}
\sum_{i=1}^{\size{X}}(n_i-\mu)^2)}$, where $\mu=\frac{1}{\size{X}}
\sum_{i=1}^{\size{X}}n_i$ (by analogy with statistical variance).  We say
that $X$ is \textit{variable} if variability of its NSS is greater than
some $T \in \mathbb{R}$.  $T$ depends on the language model $\mathcal{M}$,
and is set to 1450 in our experiments.

It is computationally infeasible to compute
$\norm{\pi\brac{X}-\pi\brac{Y}}$ for every pair $X,Y\in\mathbb{S}$
in the English language.  To validate our conjecture, we show that
when a ``variable'' $X$ and another sequence $Y$ are sampled from
a real-world English corpus and $X$ and $Y$ are not similar, it
always holds that $\norm{\pi\brac{X}-\pi\brac{Y}}>U\brac{\size{X}}$
for a large $U\brac{\size{X}}$.  Critically, $U$ depends only on the
sequence length but not the size or domain of the corpus from which
$X$ is drawn.  Furthermore, this holds for any $Y$, variable or not.
This implies that there are no other fingerprints within the $U$-radius
ball of $\pi\brac{X}$.

\paragraphbe{Generating NSS.} 
\label{sec:expsetupfp}
We downloaded 5 ``subreddit'' archives from Convokit~\cite{convokit}
that have the fewest common users: asoiaf, india, OkCupid,
electronic\_cigarette, and Random\_Acts\_of\_Amazon.  We also downloaded
the sports subreddit that has more active users and posts.  We then
aggregated each user's posts into longer sequences (up to 3000 words)
by concatenating them in chronological order.

To simulate auto-completion running in the background while a text
sequence is being typed, we invoke the Hugging Face Transformers
language generator (\texttt{run\_generation.py}) to drive a
GPT-2~\cite{radford2019language} language model (\texttt{gpt2-small})
and output one word for every prefix.  We use nucleus size with $q=0.9$.
To reduce computational complexity, we modified the script to save the
encoder's hidden state for every prefix, so it is necessary to decode
only one additional word for the next prefix.

\paragraphbe{Many sequences are variable.}
The fraction of variable sequences depends on the domain: 43.6\%
for the OkCupid dataset, 62.8\% for asoiaf, 77.6\% for india, 71.6\%
for electronic\_cigarette, 41.6\% for Random\_Acts\_Of\_Amazon,
42\% for sports.  Variability seems to be strongly and inversely
correlated with the average post length, i.e., short posts result in
high variability.  The average post length is 50.3 for OkCupid dataset,
35.8 for asoiaf, 38.0 for india, 42.9 for electronic\_cigarette, 49.2
for Random\_Acts\_Of\_Amazon, 60.6 for sports.  We conjecture that the
(re-initialized) language model is more ``uncertain'' about the next
word at the beginning of posts, and large nucleus sizes correspond to
high variability.  Figure~\ref{fig:fingerprint} illustrates this effect.

We conclude that, even when posts are relatively long (e.g., sports),
many (>40\%) sequences are variable.  This fraction may be lower when
individual texts are much longer, e.g., blog posts.

\begin{figure}[h]
\vspace{-2.1em}
\centering
\subfloat[width=0.33\textwidth][NSS pairwise distances for 500 random 2700-word
sequences (OkCupid).
\label{fig:fitted-gaussian}]{
\includegraphics[width=0.33\textwidth]{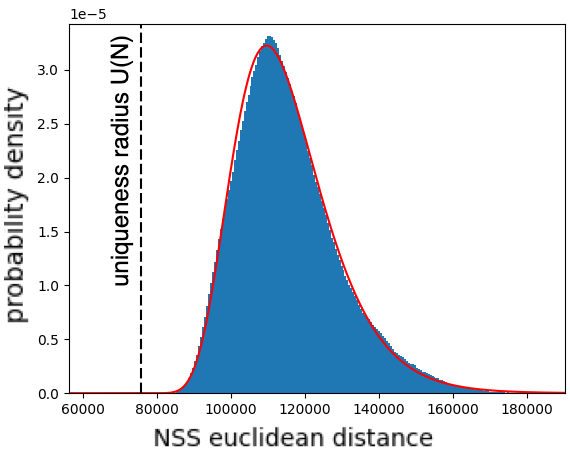}
}\hfil
\subfloat[width=0.34\textwidth][Lowest NSS distances for different dataset sizes (sports).
\label{fig:dssizes}]{
\label{fig:ball-size-subset}
\begin{tikzpicture}
\begin{axis}[
    width = 0.37\textwidth,
    legend pos = south east,
    legend columns=2,
    legend style={at={(1,0)}, nodes={scale=0.8}, inner sep=2pt},
    scaled ticks=true,
    tick label style={font=\tiny},
    label style={font=\tiny},
    xtick={500, 1000, 1500, 2000, 2500},
    ytick={20000, 30000, 40000, 50000, 60000, 70000, 80000, 90000},
    xlabel=sequence length,
    x tick label style={rotate=90,anchor=east,
        /pgf/number format/.cd,
         set thousands separator={}},
    ylabel={\tiny lowest NSS distances},
    y label style={at={(axis description cs:0.3,.5)},anchor=south},
    grid = major,
    legend cell align={left}
]

\addlegendentry{\hspace{-.6cm}\textbf{dataset size}}
\addlegendentry{4000}
\addlegendentry{3000}
\addlegendentry{2000}
\addlegendentry{1000}
\addlegendentry{500}

\addlegendimage{empty legend}

\addplot[
    color=blue,
    only marks,
    mark=o]
table[x=sequence_length, y=4000, col sep=comma]
{figures/plot_data/reddit_distance.dat};
\addplot[
    color=red,
    only marks,
    mark=triangle]
table[x=sequence_length, y=3000, col sep=comma]
{figures/plot_data/reddit_distance.dat};
\addplot[
    color=purple,
    only marks,
    mark=square]
table[x=sequence_length, y=2000, col sep=comma]
{figures/plot_data/reddit_distance.dat};
\addplot[
    color=orange,
    only marks,
    mark=diamond]
table[x=sequence_length, y=1000, col sep=comma]
{figures/plot_data/reddit_distance.dat};
\addplot[
    color=green,
    only marks,
    mark=pentagon]
table[x=sequence_length, y=500, col sep=comma]
{figures/plot_data/reddit_distance.dat};
\end{axis}

\end{tikzpicture}
}\hfil
\subfloat[width=0.35\textwidth][Lowest NSS distances for
different domains.\label{fig:ball-size-datasets}]{
\begin{tikzpicture}
\begin{axis}[
    width = 0.37\textwidth,
    legend pos = south east,
    legend columns=2,
    legend style={at={(1,0)}, cells={align=left}, nodes={scale=0.75}, inner sep=1pt},
    legend cell align={left},
    scaled ticks=true,
    tick label style={font=\tiny},
    label style={font=\tiny},
    xtick={500, 1000, 1500, 2000, 2500},
    x tick label style={rotate=90,anchor=east,
        /pgf/number format/.cd,
         set thousands separator={}},
    ytick={20000, 30000, 40000, 50000, 60000, 70000, 80000, 90000},
    y label style={at={(axis description cs:0.3,.5)},anchor=south},
    xlabel=sequence length,
    ylabel={\tiny lowest NSS distances},
    grid = major,
]

\addlegendentry{asoiaf}
\addlegendentry{india}
\addlegendentry{OkCupid}
\addlegendentry{EC}
\addlegendentry{RAOA}
\addlegendentry{U(N)}

\addplot[
    color=blue,
    only marks,
    mark=o]
table[x=sequence_length, y=asoiaf, col sep=comma]
{figures/plot_data/subreddit_distance.dat};
\addplot[
    color=red,
    only marks,
    mark=triangle]
table[x=sequence_length, y=india, col sep=comma]
{figures/plot_data/subreddit_distance.dat};
\addplot[
    color=purple,
    only marks,
    mark=square]
table[x=sequence_length, y=OkCupid, col sep=comma]
{figures/plot_data/subreddit_distance.dat};
\addplot[
    color=orange,
    only marks,
    mark=diamond]
table[x=sequence_length, y=electronic_cigarette, col sep=comma]
{figures/plot_data/subreddit_distance.dat};
\addplot[
    color=green,
    only marks,
    mark=pentagon]
table[x=sequence_length, y=Random_Acts_Of_Amazon, col sep=comma]
{figures/plot_data/subreddit_distance.dat};
\addplot[
    color=black,
    only marks,
    mark=*]
table[x=x, y={U(N)}, col sep=comma]
{figures/plot_data/unique_radius.dat};
\end{axis}
\end{tikzpicture}
}
\caption{Nucleus size series (NSS) are fingerprints.\label{fig:fingerprints}}
\vspace{-1em}
\end{figure}
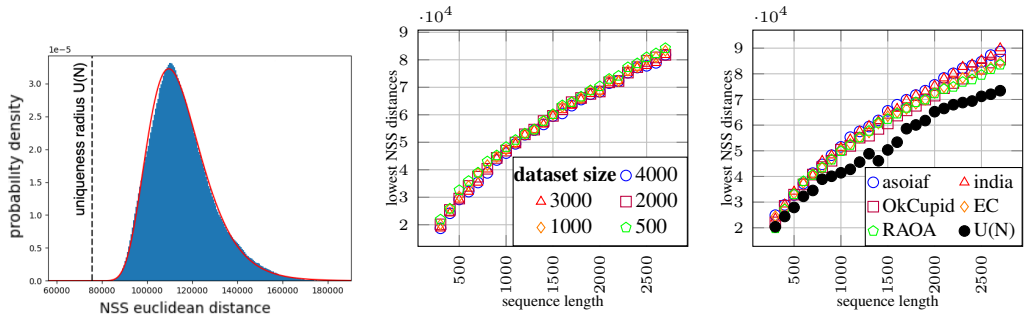

\paragraphbe{NSS of a variable sequence is unique.}
We measure pairwise Euclidean distances between the NSS of variable
sequences and the NSS of other (not necessarily variable) sequences.
Figure~\ref{fig:fitted-gaussian} shows the histogram (smoothed
by averaging over a 10-bucket window) for 500 randomly chosen
2700-word sequences from the OkCupid dataset.  Sample density decreases
exponentially with distance from the density peak, which is around 105k.
Because we omit the pairs where neither NSS is variable, this effect is
asymmetric: density decreases slower above the peak than below the peak.
Exponential decay to the left ensures that the lowest values observed
in practice are never too far from the peak.

To verify this on a larger scale, we confirmed that the lowest pairwise
distance between a variable NSS and any other NSS is consistent
regardless of the dataset size (Figure~\ref{fig:dssizes}) or domain
(Figure~\ref{fig:ball-size-datasets}).

\vspace{-2em}
\begin{minipage}{0.47\textwidth}
\begin{algorithm}[H]
\caption{Nucleus sampling~\cite{writewithtransformer}}
\scriptsize
\label{alg:nucleus}
\begin{algorithmic}[1]
\Procedure{sample\_sequence}{$\mathcal{M}, p, X$}
\State $logits \gets \mathcal{M}\brac{X}$ // get logits
\State $logits \gets \Call{top\_p\_filtering}{logits,p}$
\State $next \gets \Call{multinomial\_sample}{logits}$
\State \textbf{return} $next$
\EndProcedure
\\
\Procedure{top\_p\_filtering}{$logits,p$}
\State $sorted\_logits, indices \gets${\tiny $\Call{descend\_argsort}{logits}$}
\State $cum\_probs \gets ${\tiny $\Call{cum\_sum}{\Call{softmax}{sorted\_logits}}$}
\State $not\_in\_p \gets [\;]$
\For{$i \in \{1...\Call{len}{logits}\}$}
    \If{$cum\_probs[i] > p$}
            \State $not\_in\_p.\Call{append}{indices[i]}$
    \EndIf
\EndFor
\For{$i \in not\_in\_p$} {\color{red}$\ \ \ \ \leftarrow$ number of iterations}
    \State $logits[i] \gets -\infty$ \ \ \ \ \hspace{1em}  {\color{red}corresponds to nucleus size}
\EndFor  {\color{red}nucleus size}
\State \textbf{return} $logits$%
\EndProcedure
\end{algorithmic}
\end{algorithm}
\end{minipage}
\begin{minipage}{0.51\textwidth}
\begin{figure}[H]
\begin{tikzpicture}
\clip (-0.7,-0.5) rectangle (6.8, 5.5);
\begin{axis}[
    width = 1.15\textwidth,
    height = 0.9\textwidth,
    legend pos = north east,
    legend style={at={(1,1)},nodes={scale=0.5}, cells={align=left},},
    scaled ticks=true,
    xlabel=word index,
    ytick={0,2000,4000,6000,8000,10000},
    ymax=13000,
    x label style={at={(axis description cs:0.5,0.07)},anchor=north},
    y label style={at={(axis description cs:0.13,.5)},anchor=south},
    ylabel=nucleus size,
    label style={font=\tiny},
    tick label style={font=\tiny},
    grid = major
]

\addplot[
    color=blue,
    mark size = 1pt]
table[x=x, y=s, col sep=comma]
{figures/plot_data/fingerprint.dat};

\draw [thick,dashed] (210,-100) -- (210,10000);

\node[draw,align=left,font=\tiny,fill=white] at (70,117) {Post1: This contest is closed. Make me \\laugh with a rhyming phrase like “elf \\on a shelf” or “Obama on a llama"};
\node[draw,align=left,font=\tiny,fill=white] at (272,117) {Post2: One person at\\random at the end\\of the contest wins!};

\end{axis}
\end{tikzpicture}
\caption{Concatenated NSS of 2 posts.\label{fig:fingerprint}}
\end{figure}
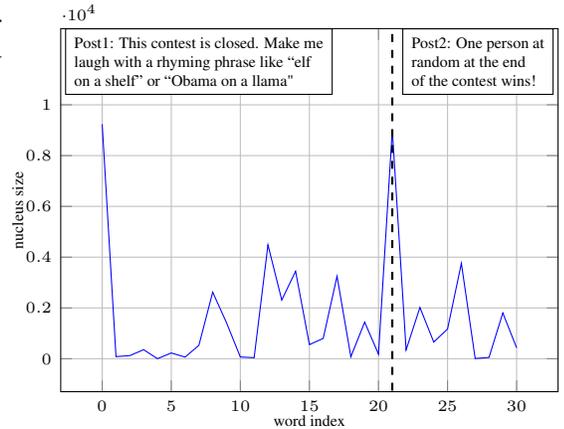
\end{minipage}\hfil

\paragraphbe{NSS of a variable sequence is a fingerprint.}
To formally satisfy our definition of a fingerprint, NSS of variable
sequences must have a uniqueness radius that depends on $N$. To
show this for a given $N$, we take the dataset with the lowest
average variability and fit a log-normal distribution, which is
the best fit among 90 distributions~\cite{fitdist}, to its sample
histogram, as shown in Figure~\ref{fig:fitted-gaussian}. We chose
$U(N)$ such that, on our fitted distribution, the probability to
sample an element lower than $U(N)$ is $\epsilon\equiv 10^{-18}$, which we consider
negligible. Figure~\ref{fig:ball-size-datasets} shows $U(N)$ for various
$N$.

\subsection{Execution path of nucleus sampling reveals nucleus sizes}
\label{sec:algovuln}

Algorithm \ref{alg:nucleus} shows the pseudo-code of nucleus sampling.
After obtaining the probability of each possible next token from the
language model, it calls \Call{top\_p\_filtering}{}, which sorts and
sums up the probabilities.  It then selects the tokens whose cumulative
probability is outside $top\_p$ and sets the corresponding logits to
$-\infty$, i.e., removes these tokens.  If an adversary can infer the
number of loop iterations in line 14, he can learn the number of tokens
removed from the vocabulary and thus the nucleus size, which is equal
to the vocabulary size minus the number of removed tokens.

Auto-completion exposes not just the nucleus size at each step, but also
the number of completed words before it stops due to low probability
or end-of-sequence token.  The series of these numbers, in addition to
nucleus sizes, may be an even stronger fingerprint, but we leave this
to future work.

\section{Attack overview}

\subsection{Threat model}

Consider a user who types text into a program that uses an auto-completion
assistant based on nucleus sampling.  At the same time, an attacker
is running a concurrent, low-privilege process on the user's machine,
(e.g., inside another application).  Memory isolation and sandboxing
ensure that the attacker's process cannot directly access the user's
keyboard entries, nor the resulting text.

The attacker's goal is to infer the nucleus size series,
which are revealed through the loop iteration count (see
Section~\ref{sec:algovuln}), via any available side channel (see
Section~\ref{sec:microattacks}).  We assume that the attacker has access
to the victim's language-model implementation.  This is plausible for
popular, publicly released code as GPT-2 and Hugging Face.  If the
victim is using an off-the-shelf auto-completion assistant as part of
a commercial software package, the same code is likely available to
potential attackers.  Therefore, for any candidate text, the attacker
can (re)produce the corresponding nucleus size series by re-running the
model on the prefixes of this text.

We also assume that the attacker's measurement of the side channel
is ``aligned,'' i.e., the attacker can tell when the language model
is queried to auto-complete a prefix (inferring this is relatively
straightforward\textemdash see Section~\ref{sec:alignment}).  The
measurement can be imprecise, but we show that the error is bounded
(see Appendix~\ref{sec:measurement} in supplementary materials).

\begin{figure}[h]
\centering
\includegraphics[scale=0.25]{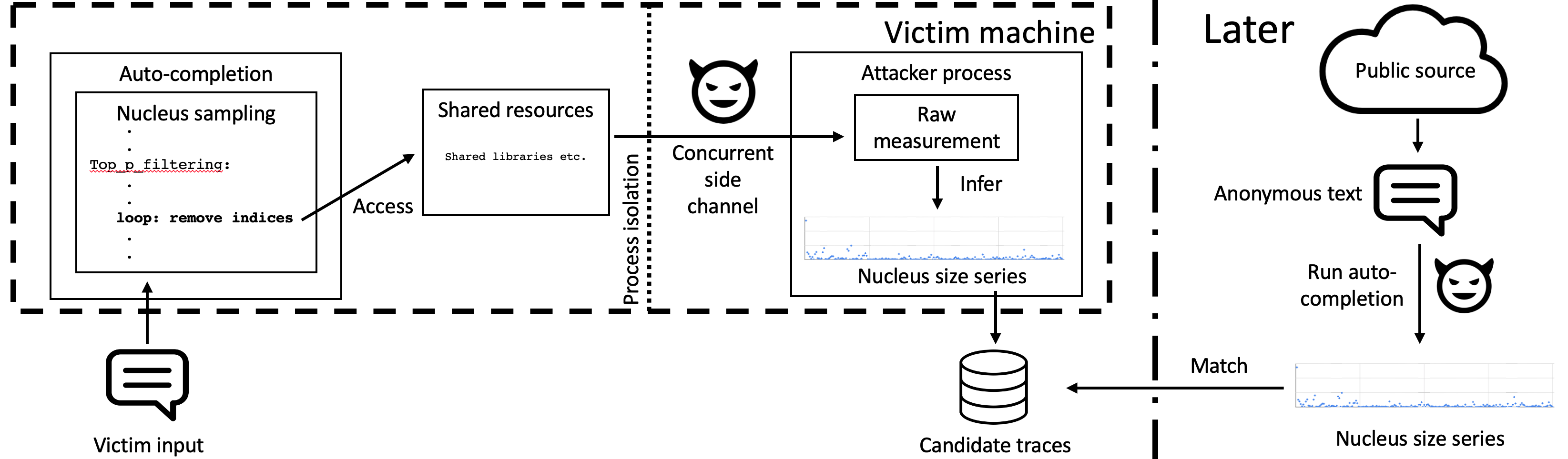}
\centering
\caption{Attack overview.}
\label{fig:attack-model}
\end{figure}

One application of this attack is de-anonymization (see
Figure~\ref{fig:attack-model}).  Consider a user who anonymously publishes
some text on Reddit, Twitter, or a blog.
In the \emph{online phase} of the attack, while the user is typing,
the attacker collects a trace by measuring the available side channel.
The attacker stores all traces, along with user identifiers such as the
IP address of the machine where the trace was collected.  In a later,
\emph{offline phase}, the attacker obtains anonymously published texts
and attempts to match them against the collected traces.

\subsection{Matching an anonymous text to a side-channel trace}
\label{sec:matching}

\begin{wrapfigure}{r}{0.6\textwidth}
\centering
\includegraphics[width=0.5\textwidth]{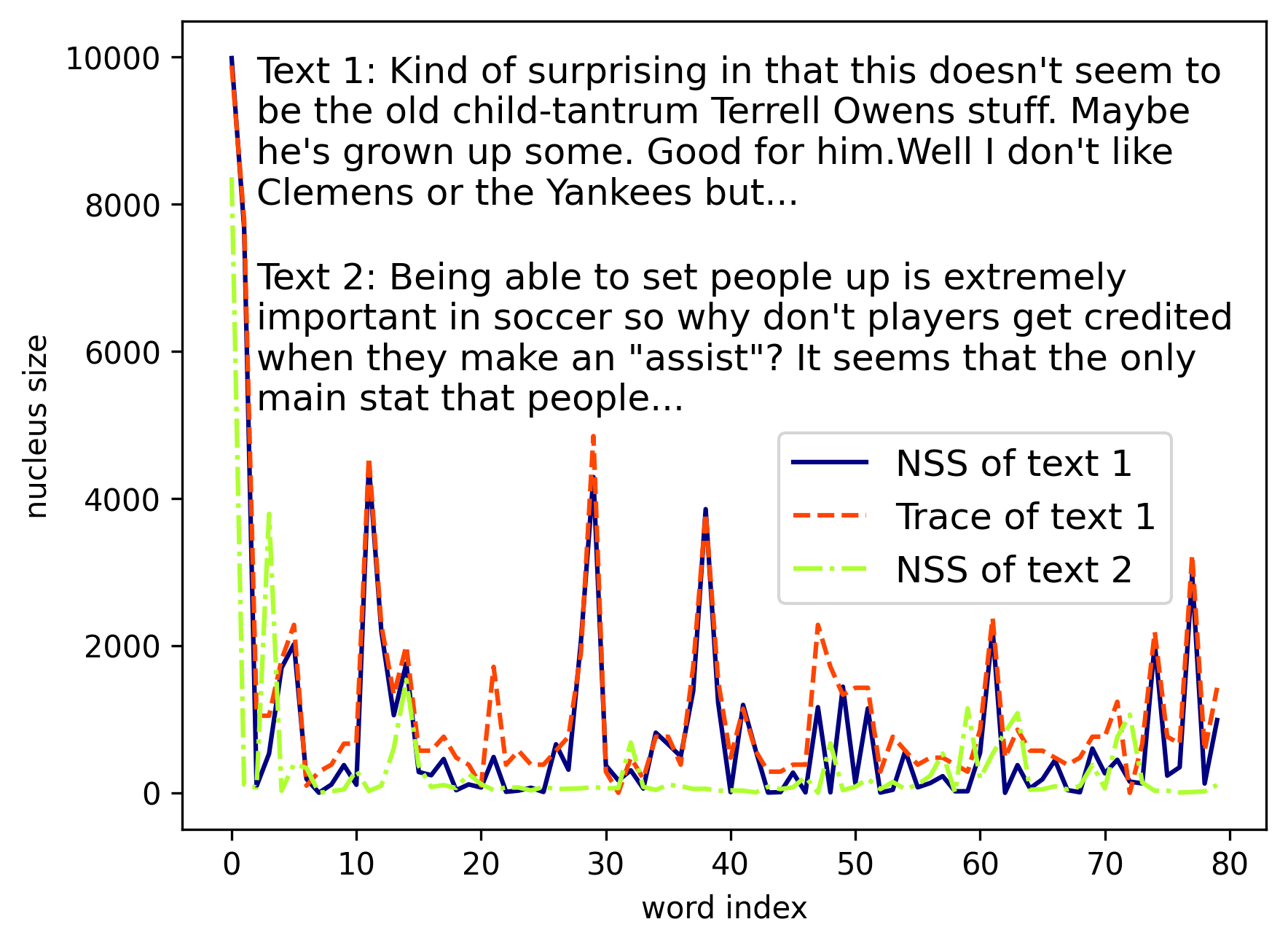}
\caption{The trace matches its text's NSS, and does not match to another text's NSS.
\label{fig:match}}
\end{wrapfigure}

Algorithm~\ref{alg:matching} shows how the attacker can match a
text sequence $X$ against previously collected traces.  First,
generate the nucleus size series (NSS) of $X$.  If this NSS is
sufficiently long and variable (see Section~\ref{sec:defs}),
$\Call{gentraces}{candidate\_traces, \size{X}}$ creates a list of
candidate traces whose length is equal to $\size{X} = N$.  It drops all
shorter traces and, for longer traces, considers all contiguous sub-traces
of length $N$.  Then compute the distance from every candidate trace
to the NSS of $X$.  If this distance is under some threshold $\tau_{N}$
that depends on $N$, declare a successful match.  Figure~\ref{fig:match}
illustrates how the trace ``matches'' the fingerprint of the correct
text but not those of other texts.

\paragraphbe{Choosing $\tau_{N}$ to ensure no false positives.}
Let $\mathbb{V}\subseteq\mathbb{S}$ the set of texts whose NSS is
variable, $X \in \mathbb{V}$ a variable text, and $Y\in \mathbb{S}$
any text s.t.\ $X,Y$ have the same length $N$ but are not similar.
Let $t$ be a trace measured while the user was typing $Y$.  We want
to avoid false positives, i.e., $Y$'s trace mistakenly matched to $X$:
$\norm{\ideal_{\mathcal{M},q}\brac{X} - t} < \tau_{N}$. Let $T^Y$ be the
probability distribution of traces measured while the user is typing $Y$,
and let $d(N)$ be the bound on the attacker's measurement error:
\begin{equation}
\label{eq:fp1}
Pr_{t\leftarrow T^{Y}}[\norm{\ideal_{\mathcal{M},q}\brac{Y} - t} < d(N)]
\geq 1 - \epsilon\hspace{2ex}\textrm{for some small $\epsilon$}
\end{equation}

From Section \ref{sec:emp}, we have that, for uniformly sampled $X$ and $Y$,
\begin{equation}
\label{eq:fp2}
Pr_{X,Y\overset{U}{\leftarrow} \mathbb{V}\times \mathbb{S}}\sqbrac{
\norm{\ideal_{\mathcal{M},q}\brac{Y} - \ideal_{\mathcal{M},q}\brac{X}} > U(N)} \geq 1 - \epsilon
\end{equation}

For any $t,X,Y$ such that the events in Equations~\ref{eq:fp1} and~\ref{eq:fp2}
hold, the distance from $t$ to $X$'s fingerprint is bound by the
triangle inequality: $\norm{\ideal_{\mathcal{M},q}\brac{X} - t} \geq
U(N)-d(N)$\textemdash see Figure~\ref{fig:bound}.  By setting the
threshold $\tau_{N}\gets U(N)-d(N)$, we guarantee that for
random $X\in\mathbb{V}$ and $Y\in\mathbb{S}$, the probability of a false positive
where $\norm{\ideal_{\mathcal{M},q}\brac{X} - t} < \tau_{N}$ is at most
$2\epsilon$ (by union bound).

\begin{minipage}{0.48\textwidth}
\begin{algorithm}[H]
\caption{Matching a sequence to trace}
\footnotesize
\label{alg:matching}
\begin{algorithmic}[1]
\Procedure{matching}{$X, candidates$}
\State $\ideal_{\mathcal{M},q}\brac{X} \gets \Call{find\_nss}{\mathcal{M},q,X}$
\If{$\ideal_{\mathcal{M},q}\brac{X}$ not \textit{variable}}
    \State \textbf{return} $not\_variable$
\EndIf
\For{$t \in \Call{gentraces}{candidates, \size{X}}$}
    \If{$\norm{\ideal_{\mathcal{M},q}\brac{X} - t} < \tau_{\size{t}}$}
        \State \textbf{return} $t$
    \EndIf
\EndFor
\State \textbf{return}  $no\_match$
\EndProcedure
\end{algorithmic}
\end{algorithm}
\end{minipage}
\hfill
\begin{minipage}{0.48\textwidth}
\begin{figure}[H]
\centering
\includegraphics[width=\textwidth]{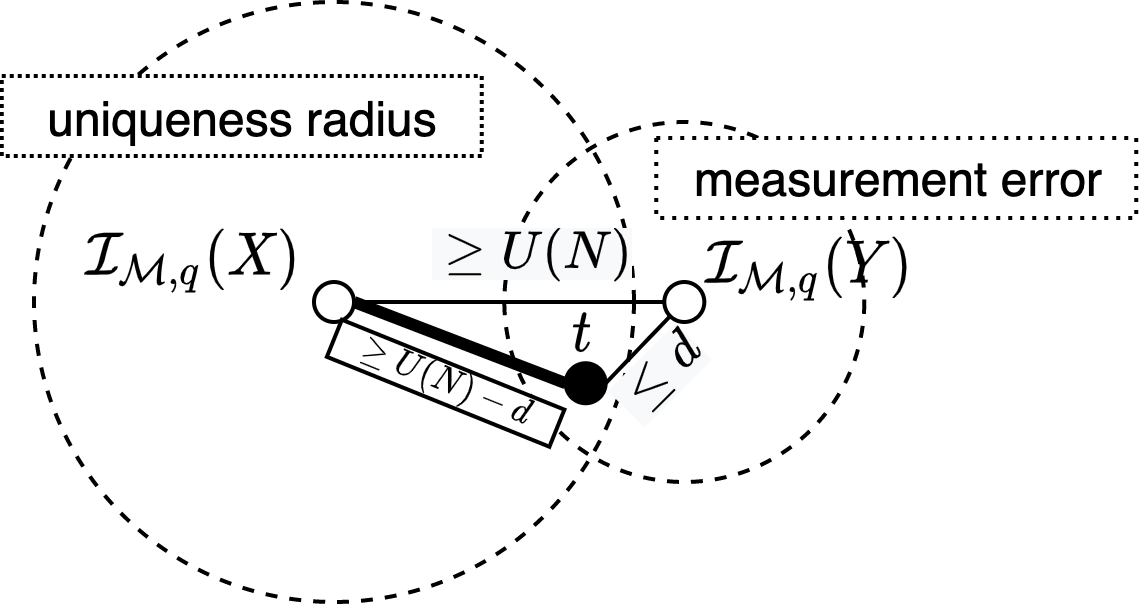}
\caption{Bound on measurement error and uniqueness radius ensures no
false positives.}
\label{fig:bound}
\end{figure}
\end{minipage}

\vspace{-1em}
\paragraphbe{Beyond distance-based matching.}
Our matching algorithm is simple, very conservative, and amenable to
theoretical analysis.  A real-world attacker who is not interested
in provable guarantees could use much more sophisticated methods.  Convolutional neural networks often outperform distance-based
methods~\cite{sirinam2018deep, schuster2017beauty}, especially with
noisy measurements~\cite{schuster2017beauty}.  For our task, these
methods are likely to be effective even when using a very noisy side
channel where $d(N)$ is higher than $U(N)$.  Empirically demonstrating
their precision for time-series fingerprint matching with an extremely
low base rate~\cite{axelsson2000base} would take many billions of
measurements, however.

\section{Fingerprinting via a cache side channel}

Our proof of concept uses a Flush+Reload attack (see
Section~\ref{sec:cacheattacks1}).  For this attack, we show that the
noise $d(N)$ of the attacker's measurements of nucleus sizes is much
smaller than the uniqueness radius $U(N)$ of nucleus size series.
This attack thus has high recall and \textbf{no false positives} (see
Section~\ref{sec:matching}).  We also demonstrate that \textbf{uniqueness
radius grows faster than measurement noise} as a function of the sequence
length $N$.  Therefore, even for noisier measurements from a different
side channel, machine, or software setup, we expect that there exists
an $N$ such that $U(N)>>d(N)$.

\subsection{Experimental setup}
\label{sec:machine-spec}

\begin{figure}[h]
\vspace{-0.5em}
\centering
\includegraphics[width=0.8\textwidth]{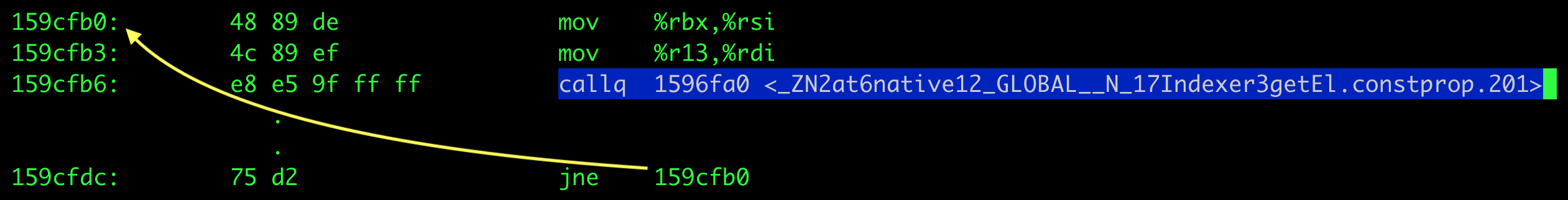}
\caption{Assembly code of the loop in lines 14-15
of Algorithm~\ref{alg:nucleus}, implemented in
\texttt{libtorch.so}.\label{fig:assembly}}
\end{figure}

Our ``victim'' uses an auto-completion app based on Hugging
Face's PyTorch code driving a GPT-2-small language model, as in
Section~\ref{sec:expsetupfp}.  We used Hugging Face~\cite{hfgithub}
and Pytorch~\cite{ptgithub} code versions from, respectively,
7/18/2019
and
7/22/2019.
The victim and attacker run as (isolated) processes on the same core of
an 8-core, Intel Xeon E5-1660 v4 CPU.  If PyTorch is installed on the
machine, the \texttt{libtorch.so} shared object (SO) is in a public,
world-readable directory.  The victim loads this SO into their process.
The attacker loads the same SO, thus the SO's physical memory addresses
are mapped into the virtual process space of both the attacker and
the victim (operating systems have a copy-on-write policy for SOs
in physical memory).  The attacker uses Flush+Reload to monitor the
first instruction of a function called within the loop, as shown in
Figure~\ref{fig:assembly}.

\vspace{-0.5em}

\label{sec:alignment}

To determine when typing starts or ends, the attacker can use any side
channel from Section~\ref{sec:microattacks} to probe the auto-completion
application or shared libraries.  For segmenting the trace into prefixes,
the attacker can use CPU timestamps for each Flush+Reload hit to
identify the gaps.  Measured traces must be processed to remove noise
and outliers\textemdash see Appendix~\ref{sec:preproc} in supplementary
materials.

Human editing such as deleting or rewriting ``pollutes'' the measured
trace with nucleus sizes corresponding to deleted subsequences.  This may
cause false negatives, but not false positives.  The attacker can try
to guess which trace chunks originate from edits and remove them before
matching.  When there are many edits, this has a nontrivial computational
cost, which can be offset by using auxiliary information from the side
channel to guide the guesses (e.g., timing, nucleus sizes, and control
flow of the code that operates the language model).  We did not evaluate
these methods because human editing is difficult to simulate at scale,
and leave them for future work.

\begin{figure}[h]
\centering

\subfloat[width=0.31\textwidth][Uniqueness radius vs.\ upper\\ bound of
measurement error.\label{fig:un-vs-d}]{
\begin{tikzpicture}
\begin{axis}[
    width = 0.38\textwidth,
    legend pos = north west,
    legend style={nodes={scale=0.8}},
    scaled ticks=true,
    tick label style={font=\tiny},
    xtick={500, 1000, 1500, 2000, 2500},
    x tick label style={rotate=90,anchor=east,
        /pgf/number format/.cd,
         set thousands separator={}},
    ytick={20000, 30000, 40000, 50000, 60000, 70000, 80000, 90000},
    xlabel=sequence length,
    ylabel={distances},
    y label style={at={(axis description cs:0.28,.5)},anchor=south},
    grid = major,
    legend cell align={left}
]

\addlegendentry{U(N)}
\addlegendentry{d(N)}

\addplot[
    color=blue,
    only marks,
    mark=o]
table[x=N, y=U(N), col sep=comma]
{figures/plot_data/recall.dat};
\addplot[
    color=red,
    only marks,
    mark=triangle]
table[x=N, y=d, col sep=comma]
{figures/plot_data/recall.dat};
\end{axis}

\end{tikzpicture}
}\hfil
\subfloat[width=0.31\textwidth][Gap between uniqueness\\ radius and error vs.\ 
median error.\label{fig:median}]{

\begin{tikzpicture}
\begin{axis}[
    width = 0.38\textwidth,
    legend pos = north west,
    legend style={nodes={scale=0.8}},
    scaled ticks=true,
    tick label style={font=\tiny},
    xtick={500, 1000, 1500, 2000, 2500},
    x tick label style={rotate=90,anchor=east,
        /pgf/number format/.cd,
         set thousands separator={}},
    ytick={20000, 30000, 40000, 50000, 60000, 70000, 80000, 90000},
    xlabel=sequence length,
    ylabel={distances},
    y label style={at={(axis description cs:0.28,.5)},anchor=south},
    grid = major,
    legend cell align={left}
]

\addlegendentry{U(N)-d(N)}
\addlegendentry{error median}

\addplot[
    color=blue,
    only marks,
    mark=o]
table[x=N, y=U(N)-d, col sep=comma]
{figures/plot_data/median.dat};
\addplot[
    color=red,
    only marks,
    mark=triangle]
table[x=N, y=error median, col sep=comma]
{figures/plot_data/median.dat};
\end{axis}

\end{tikzpicture}
}\hfil
\subfloat[width=0.31\textwidth][Attack recall for non-noisy traces (after processing).\label{fig:recall}]{
\begin{tikzpicture}
\begin{axis}[
    width = 0.38\textwidth,
    legend pos = north west,
    legend style={nodes={scale=0.5}},
    scaled ticks=true,
    tick label style={font=\tiny},
    label style={font=\footnotesize},
    xtick={500, 1000, 1500, 2000, 2500},
    x tick label style={rotate=90,anchor=east,
        /pgf/number format/.cd,
         set thousands separator={}},
    ytick={20, 40, 60, 80, 100},
    xlabel=sequence length,
    ylabel={recall (\%)},
    y label style={at={(axis description cs:0.25,.5)},anchor=south},
    grid = major,
    legend cell align={left}
]

\addplot[
    color=blue,
    only marks,
    mark=o]
table[x=N, y=recall, col sep=comma]
{figures/plot_data/recall.dat};
\end{axis}

\end{tikzpicture}
}
\caption{Measurement error and attack recall.\label{fig:error-and-recall}}
\end{figure}
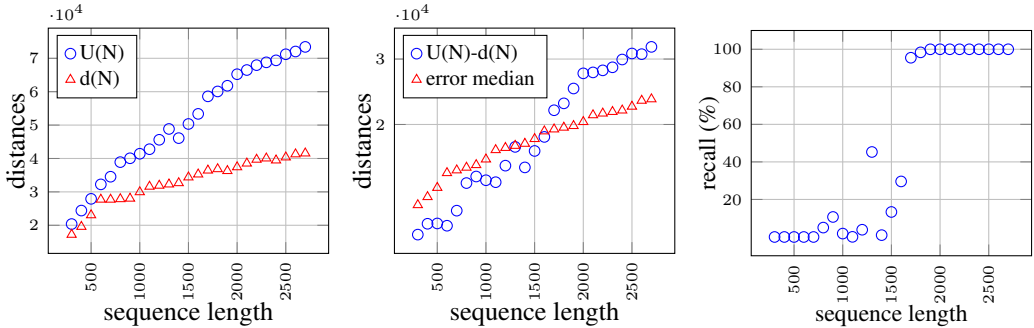
\vspace{-0.5em}

\subsection{Measurement error and attack recall}

The analysis in Section \ref{sec:matching} assumes that the side-channel
measurement error is bounded by some $d(N)$.  After measuring 1566 traces
from the reddit-sports dataset and removing noisy traces, we fit a normal
distribution and set $d(N)$ to 10 standard deviations above the mean.  The
fit is tight but not perfect\textemdash see Appendix~\ref{sec:preproc}.
Figure \ref{fig:un-vs-d} shows $d(N)$ as a function of $N$, and how the
uniqueness radius $U(N)$ diverges from $d(N)$.

This illustrates the fundamental characteristic that enables our attack:
\textbf{pairwise distances between fingerprints grow faster as a function
of sequence length $N$ than the attacker's measurement error}.  Therefore,
if a sequence is long enough, the attacker can match the fingerprint
without false positives.  This property can hold for any side channel,
not just Flush+Reload.  The key requirement is that the (squared) error of
a single measurement is, on average, smaller than the (squared) difference
between the nucleus sizes in the same position of different sequences.

In Algorithm \ref{alg:matching}, we set the threshold $\tau_{N}$
to $U(N)-d(N)$, so that the recall, after noisy trace filtering, is
equal to the probability that measurement error is below $U(N) - d(N)$.
Figure~\ref{fig:recall} shows recall for different $N$: when $N \geq
1900$, recall is greater than 99\%.  When accounting for the 6\% of
traces that were filtered out as too noisy (Appendix~\ref{sec:preproc}),
this is equivalent to recall >93\%.

\subsection{Case studies}

We show that users of real-world anonymous forums would have been
vulnerable if they had used auto-completion based on nucleus sampling.

\paragraphbe{Silk Road forum.}
We used an archive of Silk Road forum posts~\cite{silkroad} created by one
of the participants in October 2013, after the Silk Road marketplace was
shut down but before the forums were taken offline.  For each of 41 users
who had at least 2700 words in their posts, we concatenated their posts in
chronological order into a single sequence and generated the corresponding
NSS fingerprints.  An individual post has 50.6 words on average.

We simulated the auto-completion process for each user's sequence
using the Hugging Face Transformers language generator and applied our
proof-of-concept attack from Section~\ref{sec:expsetupfp}.  In reality,
posts may be separated by unrelated typing, but (a) it is relatively
straightforward to identify the current application via techniques
from Section~\ref{sec:microattacks}, and (b) the attacker knows
when the typing begins and ends (see Section~\ref{sec:alignment}).
To ensure even stronger isolation, we ran the attack process in an
AppArmor~\cite{apparmor} sandbox (by default, it still lets the attacker
read PyTorch shared objects).  These experiments were done on the same
machine as in Section~\ref{sec:machine-spec}.

We truncated all traces to $N$=2700 and filtered out NSS that
are not sufficiently variable.  This left 18 users out of 41.
For each of them, we computed the measurement error of the attack,
i.e., the distance between the measured trace and NSS\textemdash see
Appendix~\ref{sec:silkroaddata} in supplementary materials.  In all cases,
the error is less than $U(N) - d(N)$, thus the attack would have been
able to correctly de-anonymize these 18 users with no false positives
or false negatives.

\paragraphbe{Ubuntu Chat.}
We selected the 200 most active users from the Ubuntu Chat
corpus~\cite{ubuntu} and followed the same procedure as above to generate
the NSS fingerprints of their posts.  The average post length is 9.5
words.  Because posts are short, the sequences of all selected users are
sufficiently variable.  Filtering out noisy traces and (for technical
reasons) 2 users with irregular characters in their usernames left 186
traces.  For all of them, the error was less than $d(N)$, so there were
no false positives, and for all except one, the error was less than $U(N)
- d(N)$, so they were identified correctly.  The overall recall is 93.4\%.

\section{Mitigation}

To replace Algorithm~\ref{alg:nucleus}, we suggest
Algorithm~\ref{alg:mitigation} which follows two standard guidelines for
cryptographic code.  First, it avoids data-dependent memory accesses
at a granularity coarser than a cache line~\cite{ches-2011-29770,
gueron2012efficient}, thus an attacker cannot mount a Flush+Reload
attack to count how many times a code line executes.  Whereas
Algorithm~\ref{alg:nucleus} iterates over indices $i$ where $cum\_probs[i]
> p$ (testing if $i$ is within the $p$-nucleus) and assigns $-\infty$,
Algorithm~\ref{alg:mitigation} entirely avoids control flows that
depend on the condition $cum\_probs[i] > p$.\footnote{To this end,
we use the expression $\Call{float}{cum\_probs[i] > p}\cdot
\mathrm{MAX\_FLOAT} \cdot 2$ in Line 6, which resolves to either
$\infty$ if $cum\_probs[i] > p$, or 0 otherwise. The multiplication
by 2 invokes a float overflow in the case of $cum\_probs[i] >
p$, which resolves to $\infty$, as per IEEE 754 floating point
arithmetic standard~\cite{kahan1996ieee}.} Second, execution time
should not be correlated with secret data~\cite{ge2018survey}.
Figures~\ref{fig:no-mitigation} and~\ref{fig:mitigation} show the
relationship between the nucleus size and execution time of the token
removal loop with and without the mitigation, indicating that our
implementation (which has a fixed number of iterations) reduces the
correlation.

The cost of our mitigation is a 1.15x average slowdown in the loop
execution time, which translates into only a 0.1\% increase in the
runtime of $\Call{sample\_sequence}{}$ (which itself accounts for a tiny
fraction of the execution time relative to the encoder/decoder passes).
When simulating auto-completion on a 2700-word sequence in the setup from
Section~\ref{sec:machine-spec}, there was a negligible, 0.3\% runtime
difference in favor of our algorithm, implying that the difference between
Algorithms~\ref{alg:nucleus} and~\ref{alg:mitigation} is dominated by
other factors, such as natural fluctuations in the CPU load.

\vspace{-1em}
\begin{minipage}{0.5\textwidth}
\begin{algorithm}[H]
\caption{Top-p filtering with a fixed number of loop iterations.}
\scriptsize
\label{alg:mitigation}
\begin{algorithmic}[1]

\Procedure{top\_p\_filtering}{$logits,p$}

\State $sorted\_logits, indices \gets${\scriptsize $\Call{argsort\_descend}{logits}$}
\State $cum\_probs \gets ${\scriptsize $\Call{cum\_sum}{\Call{softmax}{sorted\_logits}}$}
\\
\For{$i \in \{1...\Call{len}{logits}\}$}
    
	\State $z\gets \Call{float}{cum\_probs[i]>p}\cdot \mathrm{MAXFLOAT}\cdot2$
        \State $logits[indices[i]] \gets logits[indices[i]] - z$    
\EndFor
\\
\State \textbf{return} $logits$
\EndProcedure

\end{algorithmic}
\end{algorithm}
\end{minipage}\hfill
\begin{minipage}{0.5\textwidth}
\begin{figure}[H]
\centering

\subfloat[width=\textwidth][\footnotesize Nucleus size vs.\ run time (original)\label{fig:no-mitigation}]{
\includegraphics[width=\textwidth, height=0.3\textwidth]{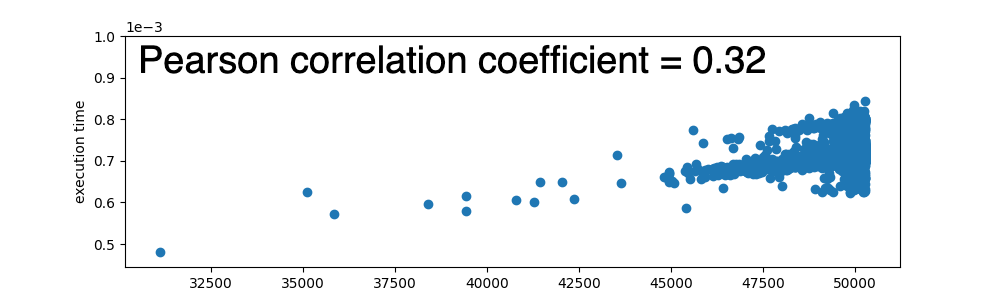}
}
\vspace{-1em}
\subfloat[width=\textwidth][\footnotesize Nucleus size vs.\ run time (mitigation)\label{fig:mitigation}]{
\includegraphics[width=\textwidth, height=0.3\textwidth]{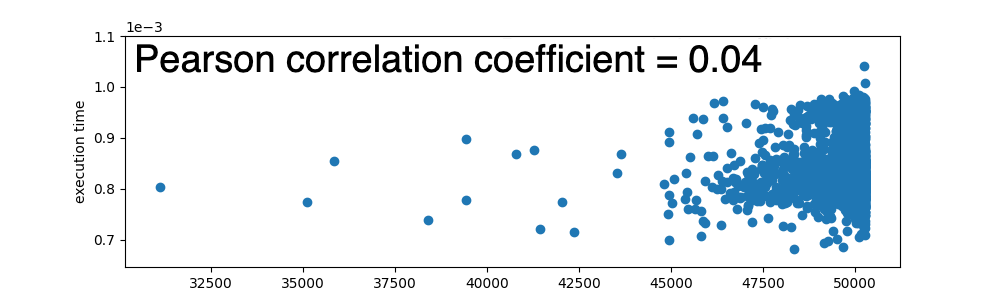}
}
\caption{Execution time of the token removal loop with and without the mitigation.}
\end{figure}
\end{minipage}

\vspace{-0.15em}

No implementation is immune to side channels, however.  Address
bits within a cache line could still leak through the cache on
certain processors~\cite{yarom2017cachebleed}.  Even without
input-dependent paths, loop runtimes may still slightly depend on
the input due to value-dependent execution times of floating-point
operations~\cite{andrysco2015subnormal} (an attacker must be able to
measure the loop very accurately to exploit this).  Mitigations of
these and other side-channel risks incur implementational and runtime
overheads~\cite{ge2018survey}.

We believe that our implementation strikes a good balance by substantially
increasing the gap between what side-channel attacks can achieve on
specific platforms in controlled laboratory conditions vs.\ what is
available to real-world attackers.  We argue that removing ``easy''
targets like input-dependent loops should be a minimal security standard
for core ML building blocks.

\section{Related work}

Prior work showed how to infer model architectures and
weights\textemdash but not inputs\textemdash via model
execution time~\cite{duddu2018stealing}, addresses of memory
accesses leaked by GPUs~\cite{hu2019neural} and trusted hardware
enclaves~\cite{hua2018reverse}, and or via cache~\cite{yan2018cache,
hong20200wn} and GPU~\cite{naghibijouybari2018rendered} side channels.

The only prior work on inferring model inputs required hardware
attacks, such as physically probing the power consumption of an
FPGA accelerator~\cite{wei2018know}, physically probing an external
microcontroller executing the model~\cite{batina2018csi}, or inferring
coarse information about the input's class from hardware performance
counters~\cite{alam2019secure}.  To the best of our knowledge, ours is
the first work to show the feasibility of inferring neural-network inputs
in a conventional, software-only setting, where the attacker is limited
to executing an isolated malicious application on the victim's machine.

\section{Conclusions}

We used nucleus sampling, a popular approach for text generation, as a
case study of ML systems that unwittingly leak their confidential inputs.
As our main technical contribution, we demonstrated that the series
of nucleus sizes associated with an English-language word sequence is
a fingerprint which uniquely identifies this sequence.  We showed how
a side-channel attacker can measure these fingerprints and use them to
de-anonymize anonymous text.  Finally, we explained how to mitigate this
leak by reducing input-dependent control flows in the implementations
of ML systems.

\newpage

\section*{Broader Impact}
This work will help improve security of ML code by (a) identifying a new
category of potential vulnerabilities faced by ML systems that operate
on sensitive data, and (b) explaining how to design implementations so as
to mitigate this risk.  This research will primarily benefit implementors
of ML models and, in general, increase trust in ML systems.

\section*{Acknowledgments and Funding}
This research was supported in part by NSF grants 1704296 and 1916717,
the Blavatnik Interdisciplinary Cyber Research Center (ICRC), the
generosity of Eric and Wendy Schmidt by recommendation of the Schmidt
Futures program, and a Google Faculty Research Award.  Roei Schuster is
a member of the Check Point Institute of Information Security.

\bibliographystyle{abbrvnat}
\bibliography{main}

\begin{thebibliography}{50}
\providecommand{\natexlab}[1]{#1}
\providecommand{\url}[1]{\texttt{#1}}
\expandafter\ifx\csname urlstyle\endcsname\relax
  \providecommand{\doi}[1]{doi: #1}\else
  \providecommand{\doi}{doi: \begingroup \urlstyle{rm}\Url}\fi

\bibitem[Alam and Mukhopadhyay(2019)]{alam2019secure}
M.~Alam and D.~Mukhopadhyay.
\newblock How secure are deep learning algorithms from side-channel based
  reverse engineering?
\newblock In \emph{{DAC}}, 2019.

\bibitem[Andrysco et~al.(2015)Andrysco, Kohlbrenner, Mowery, Jhala, Lerner, and
  Shacham]{andrysco2015subnormal}
M.~Andrysco, D.~Kohlbrenner, K.~Mowery, R.~Jhala, S.~Lerner, and H.~Shacham.
\newblock On subnormal floating point and abnormal timing.
\newblock In \emph{{S{\&}P}}, 2015.

\bibitem[{AppArmor}()]{apparmor}
{AppArmor}.
\newblock \url{https://gitlab.com/apparmor/apparmor/-/wikis/home}, 1998.
\newblock accessed: May 2020.

\bibitem[Axelsson(2000)]{axelsson2000base}
S.~Axelsson.
\newblock The base-rate fallacy and the difficulty of intrusion detection.
\newblock \emph{{TISSEC}}, 3\penalty0 (3):\penalty0 186--205, 2000.

\bibitem[Batina et~al.(2019)Batina, Bhasin, Jap, and Picek]{batina2018csi}
L.~Batina, S.~Bhasin, D.~Jap, and S.~Picek.
\newblock {CSI} neural network: Using side-channels to recover your artificial
  neural network information.
\newblock In \emph{{USENIX} {Security}}, 2019.

\bibitem[Brickell(2011)]{ches-2011-29770}
E.~F. Brickell.
\newblock Technologies to improve platform security.
\newblock In \emph{CHES}, 2011.

\bibitem[Cohney et~al.(2020)Cohney, Kwong, Paz, Genkin, Heninger, Ronen, and
  Yarom]{cohney2020pseudorandom}
S.~Cohney, A.~Kwong, S.~Paz, D.~Genkin, N.~Heninger, E.~Ronen, and Y.~Yarom.
\newblock Pseudorandom black swans: Cache attacks on {CTR DRBG}.
\newblock In \emph{{S{\&}P}}, 2020.

\bibitem[ConvoKit()]{convokit}
ConvoKit.
\newblock Cornell conversational analysis toolkit.
\newblock \url{https://convokit.cornell.edu/}, 2020.
\newblock accessed: June 2020.

\bibitem[Duddu et~al.(2018)Duddu, Samanta, Rao, and Balas]{duddu2018stealing}
V.~Duddu, D.~Samanta, D.~V. Rao, and V.~E. Balas.
\newblock Stealing neural networks via timing side channels.
\newblock \emph{arXiv:1812.11720}, 2018.

\bibitem[Fit distribution module/script (fitdist)()]{fitdist}
Fit distribution module/script (fitdist).
\newblock \url{https://github.com/alreich/fitdist}, 2020.
\newblock accessed: June 2020.

\bibitem[Ge et~al.(2018)Ge, Yarom, Cock, and Heiser]{ge2018survey}
Q.~Ge, Y.~Yarom, D.~Cock, and G.~Heiser.
\newblock A survey of microarchitectural timing attacks and countermeasures on
  contemporary hardware.
\newblock \emph{Journal of Cryptographic Engineering}, 8\penalty0 (1):\penalty0
  1--27, 2018.

\bibitem[Genkin et~al.(2014)Genkin, Shamir, and Tromer]{genkin2014rsa}
D.~Genkin, A.~Shamir, and E.~Tromer.
\newblock {RSA} key extraction via low-bandwidth acoustic cryptanalysis.
\newblock In \emph{CRYPTO}, 2014.

\bibitem[Genkin et~al.(2016)Genkin, Pachmanov, Pipman, and
  Tromer]{genkin2016ecdh}
D.~Genkin, L.~Pachmanov, I.~Pipman, and E.~Tromer.
\newblock {ECDH} key-extraction via low-bandwidth electromagnetic attacks on
  {PCs}.
\newblock In \emph{{CT-RSA}}, 2016.

\bibitem[Genkin et~al.(2018)Genkin, Pachmanov, Tromer, and
  Yarom]{genkin2018drive}
D.~Genkin, L.~Pachmanov, E.~Tromer, and Y.~Yarom.
\newblock Drive-by key-extraction cache attacks from portable code.
\newblock In \emph{ACNS}, 2018.

\bibitem[Gras et~al.(2018)Gras, Razavi, Bos, and
  Giuffrida]{gras2018translation}
B.~Gras, K.~Razavi, H.~Bos, and C.~Giuffrida.
\newblock Translation leak-aside buffer: Defeating cache side-channel
  protections with {TLB} attacks.
\newblock In \emph{{USENIX Security}}, 2018.

\bibitem[Gueron(2012)]{gueron2012efficient}
S.~Gueron.
\newblock Efficient software implementations of modular exponentiation.
\newblock \emph{Journal of Cryptographic Engineering}, 2\penalty0 (1):\penalty0
  31--43, 2012.

\bibitem[Gullasch et~al.(2011)Gullasch, Bangerter, and
  Krenn]{gullasch2011cache}
D.~Gullasch, E.~Bangerter, and S.~Krenn.
\newblock Cache games--bringing access-based cache attacks on {AES} to
  practice.
\newblock In \emph{{S{\&}P}}, 2011.

\bibitem[Hashimoto et~al.(2019)Hashimoto, Zhang, and
  Liang]{hashimoto2019unifying}
T.~B. Hashimoto, H.~Zhang, and P.~Liang.
\newblock Unifying human and statistical evaluation for natural language
  generation.
\newblock In \emph{{NAACL}}, 2019.

\bibitem[Holtzman et~al.(2020)Holtzman, Buys, Forbes, and
  Choi]{holtzman2019curious}
A.~Holtzman, J.~Buys, M.~Forbes, and Y.~Choi.
\newblock The curious case of neural text degeneration.
\newblock In \emph{ICLR}, 2020.

\bibitem[Hong et~al.(2020)Hong, Davinroy, Kaya, Dachman-Soled, and
  Dumitra{\c{s}}]{hong20200wn}
S.~Hong, M.~Davinroy, Y.~Kaya, D.~Dachman-Soled, and T.~Dumitra{\c{s}}.
\newblock How to 0wn {NAS} in your spare time.
\newblock In \emph{{ICLR}}, 2020.

\bibitem[Hu et~al.(2020)Hu, Liang, Deng, Li, Xie, Ji, Ding, Liu, Sherwood, and
  Xie]{hu2019neural}
X.~Hu, L.~Liang, L.~Deng, S.~Li, X.~Xie, Y.~Ji, Y.~Ding, C.~Liu, T.~Sherwood,
  and Y.~Xie.
\newblock Neural network model extraction attacks in edge devices by hearing
  architectural hints.
\newblock In \emph{{ASPLOS}}, 2020.

\bibitem[Hua et~al.(2018)Hua, Zhang, and Suh]{hua2018reverse}
W.~Hua, Z.~Zhang, and G.~E. Suh.
\newblock Reverse engineering convolutional neural networks through
  side-channel information leaks.
\newblock In \emph{{DAC}}, 2018.

\bibitem[{Hugging Face}()]{hfgithub}
{Hugging Face}.
\newblock Transformers on github.
\newblock \url{https://github.com/huggingface/transformers}, 2020.
\newblock accessed: June 2020.

\bibitem[Hugging Face()]{writewithtransformer}
Hugging Face.
\newblock Write with {Transfomer} (demo).
\newblock \url{https://transformer.huggingface.co/}, 2020.
\newblock accessed: June 2020.

\bibitem[Hund et~al.(2013)Hund, Willems, and Holz]{hund2013practical}
R.~Hund, C.~Willems, and T.~Holz.
\newblock Practical timing side channel attacks against kernel space {ASLR}.
\newblock In \emph{{S{\&}P}}, 2013.

\bibitem[Kahan(1996)]{kahan1996ieee}
W.~Kahan.
\newblock {IEEE} standard 754 for binary floating-point arithmetic.
\newblock \emph{Lecture Notes on the Status of {IEEE}}, 754\penalty0
  (94720-1776):\penalty0 11, 1996.

\bibitem[Kocher(1996)]{kocher1996timing}
P.~C. Kocher.
\newblock Timing attacks on implementations of {Diffie-Hellman}, {RSA}, {DSS},
  and other systems.
\newblock In \emph{CRYPTO}, 1996.

\bibitem[Lipp et~al.(2016)Lipp, Gruss, Spreitzer, Maurice, and
  Mangard]{lipp2016armageddon}
M.~Lipp, D.~Gruss, R.~Spreitzer, C.~Maurice, and S.~Mangard.
\newblock {ARMageddon}: Cache attacks on mobile devices.
\newblock In \emph{{USENIX Security}}, 2016.

\bibitem[Liu et~al.(2015)Liu, Yarom, Ge, Heiser, and Lee]{liu2015last}
F.~Liu, Y.~Yarom, Q.~Ge, G.~Heiser, and R.~B. Lee.
\newblock Last-level cache side-channel attacks are practical.
\newblock In \emph{{S{\&}P}}, 2015.

\bibitem[Naghibijouybari et~al.(2018)Naghibijouybari, Neupane, Qian, and
  Abu-Ghazaleh]{naghibijouybari2018rendered}
H.~Naghibijouybari, A.~Neupane, Z.~Qian, and N.~Abu-Ghazaleh.
\newblock Rendered insecure: {GPU} side channel attacks are practical.
\newblock In \emph{{CCS}}, 2018.

\bibitem[Oren et~al.(2015)Oren, Kemerlis, Sethumadhavan, and
  Keromytis]{oren2015spy}
Y.~Oren, V.~P. Kemerlis, S.~Sethumadhavan, and A.~D. Keromytis.
\newblock The spy in the sandbox: Practical cache attacks in {JavaScript} and
  their implications.
\newblock In \emph{{CCS}}, 2015.

\bibitem[Osvik et~al.(2006)Osvik, Shamir, and Tromer]{osvik2006cache}
D.~A. Osvik, A.~Shamir, and E.~Tromer.
\newblock Cache attacks and countermeasures: the case of {AES}.
\newblock In \emph{{CT-RSA}}, 2006.

\bibitem[Percival(2005)]{percival2005cache}
C.~Percival.
\newblock Cache missing for fun and profit.
\newblock \url{https://www.daemonology.net/papers/htt.pdf}, 2005.

\bibitem[{PyTorch}()]{ptgithub}
{PyTorch}.
\newblock \url{https://github.com/pytorch/pytorch}, 2020.
\newblock accessed: June 2020.

\bibitem[Radford et~al.(2019)Radford, Wu, Child, Luan, Amodei, and
  Sutskever]{radford2019language}
A.~Radford, J.~Wu, R.~Child, D.~Luan, D.~Amodei, and I.~Sutskever.
\newblock Language models are unsupervised multitask learners.
\newblock \emph{OpenAI Blog}, 1\penalty0 (8), 2019.

\bibitem[Ristenpart et~al.(2009)Ristenpart, Tromer, Shacham, and
  Savage]{ristenpart2009hey}
T.~Ristenpart, E.~Tromer, H.~Shacham, and S.~Savage.
\newblock Hey, you, get off of my cloud: Exploring information leakage in
  third-party compute clouds.
\newblock In \emph{{CCS}}, 2009.

\bibitem[Ronen et~al.(2019)Ronen, Gillham, Genkin, Shamir, Wong, and
  Yarom]{ronen20199}
E.~Ronen, R.~Gillham, D.~Genkin, A.~Shamir, D.~Wong, and Y.~Yarom.
\newblock The 9 lives of {Bleichenbacher}'s {CAT}: New {C}ache {AT}tacks on
  {TLS} implementations.
\newblock In \emph{{S{\&}P}}, 2019.

\bibitem[Schuster et~al.(2017)Schuster, Shmatikov, and
  Tromer]{schuster2017beauty}
R.~Schuster, V.~Shmatikov, and E.~Tromer.
\newblock Beauty and the {Burst}: Remote identification of encrypted video
  streams.
\newblock In \emph{{USENIX} {Security}}, 2017.

\bibitem[Silk Road | Users()]{silkroad}
Silk Road | Users.
\newblock \url{https://antilop.cc/sr/users/}., 2020.
\newblock accessed: June 2020. \textbf{Contains offensive and harmful
  materials}.

\bibitem[Sirinam et~al.(2018)Sirinam, Imani, Juarez, and
  Wright]{sirinam2018deep}
P.~Sirinam, M.~Imani, M.~Juarez, and M.~Wright.
\newblock Deep fingerprinting: Undermining website fingerprinting defenses with
  deep learning.
\newblock In \emph{{CCS}}, 2018.

\bibitem[Ubuntu Chat Corpus()]{ubuntu}
Ubuntu Chat Corpus.
\newblock \url{https://daviduthus.org/UCC/}, 2020.
\newblock accessed: August 2020.

\bibitem[Wei et~al.(2018)Wei, Luo, Li, Liu, and Xu]{wei2018know}
L.~Wei, B.~Luo, Y.~Li, Y.~Liu, and Q.~Xu.
\newblock I know what you see: Power side-channel attack on convolutional
  neural network accelerators.
\newblock In \emph{{ACSAC}}, 2018.

\bibitem[Welleck et~al.(2020)Welleck, Kulikov, Kim, Pang, and
  Cho]{welleck2020consistency}
S.~Welleck, I.~Kulikov, J.~Kim, R.~Y. Pang, and K.~Cho.
\newblock Consistency of a recurrent language model with respect to incomplete
  decoding.
\newblock \emph{arXiv:2002.02492}, 2020.

\bibitem[Wolf et~al.(2019)Wolf, Debut, Sanh, Chaumond, Delangue, Moi, Cistac,
  Rault, Louf, Funtowicz, and Brew]{Wolf2019HuggingFacesTS}
T.~Wolf, L.~Debut, V.~Sanh, J.~Chaumond, C.~Delangue, A.~Moi, P.~Cistac,
  T.~Rault, R.~Louf, M.~Funtowicz, and J.~Brew.
\newblock {HuggingFace}'s transformers: State-of-the-art natural language
  processing.
\newblock \emph{arXiv:1910.03771}, 2019.

\bibitem[Yan et~al.(2019)Yan, Sprabery, Gopireddy, Fletcher, Campbell, and
  Torrellas]{yan2019attack}
M.~Yan, R.~Sprabery, B.~Gopireddy, C.~Fletcher, R.~Campbell, and J.~Torrellas.
\newblock Attack directories, not caches: Side channel attacks in a
  non-inclusive world.
\newblock In \emph{{S{\&}P}}, 2019.

\bibitem[Yan et~al.(2020)Yan, Fletcher, and Torrellas]{yan2018cache}
M.~Yan, C.~Fletcher, and J.~Torrellas.
\newblock Cache telepathy: Leveraging shared resource attacks to learn {DNN}
  architectures.
\newblock In \emph{{USENIX} {Security}}, 2020.

\bibitem[Yarom and Falkner(2014)]{yarom2014flush+}
Y.~Yarom and K.~Falkner.
\newblock {FLUSH+RELOAD}: A high resolution, low noise, {L3} cache side-channel
  attack.
\newblock In \emph{{USENIX Security}}, 2014.

\bibitem[Yarom et~al.(2017)Yarom, Genkin, and Heninger]{yarom2017cachebleed}
Y.~Yarom, D.~Genkin, and N.~Heninger.
\newblock {CacheBleed}: a timing attack on {OpenSSL} constant-time {RSA}.
\newblock \emph{Journal of Cryptographic Engineering}, 7\penalty0 (2):\penalty0
  99--112, 2017.

\bibitem[Zhang et~al.(2012)Zhang, Juels, Reiter, and
  Ristenpart]{zhang2012cross}
Y.~Zhang, A.~Juels, M.~K. Reiter, and T.~Ristenpart.
\newblock {Cross-VM} side channels and their use to extract private keys.
\newblock In \emph{{CCS}}, 2012.

\bibitem[Zhang et~al.(2014)Zhang, Juels, Reiter, and
  Ristenpart]{zhang2014cross}
Y.~Zhang, A.~Juels, M.~K. Reiter, and T.~Ristenpart.
\newblock Cross-tenant side-channel attacks in {PaaS} clouds.
\newblock In \emph{{CCS}}, 2014.

\end{thebibliography}
\newpage
\appendix
\section{Trace preprocessing}
\label{sec:preproc}

\paragraphbe{Challenge: loop iterations are faster than Flush+Reload.}
The attacker's goal is to infer the number of iterations of the
token removal loop (see Section~\ref{sec:algovuln}).  In the Reload
phase of the Flush+Reload attack, the attacker learns whether
the victim has accessed an address since it has been Flushed (see
Section~\ref{sec:cacheattacks1}).  A naive attack would iteratively
perform Flush+Reload and receive indications whenever the victim accesses
this address, which happens on every iteration of the target loop.

\begin{figure}[H]
\centering
\subfloat[width=0.47\textwidth][Measurement error of 1566 2700-word traces\label{fig:noisy-distribution}]{
\includegraphics[width=0.47\textwidth]{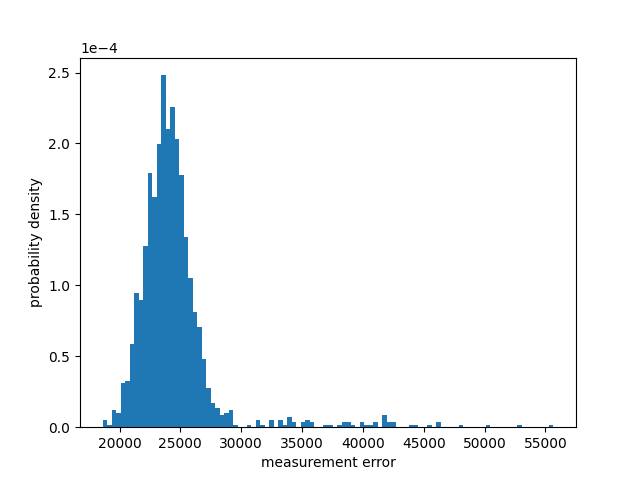}
}\hfil
\subfloat[width=0.47\textwidth][Part of a noisy trace; red circles indicate outliers\label{fig:missing-reads}]{
\includegraphics[width=0.47\textwidth]{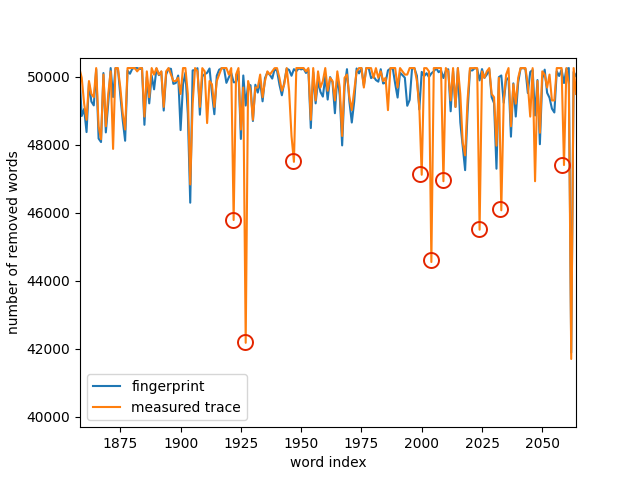}
}\hfil
\subfloat[width=0.47\textwidth][Measured number of iterations vs.\ CPU cycles: normal trace\label{fig:good-trace}]{
\includegraphics[width=0.47\textwidth]{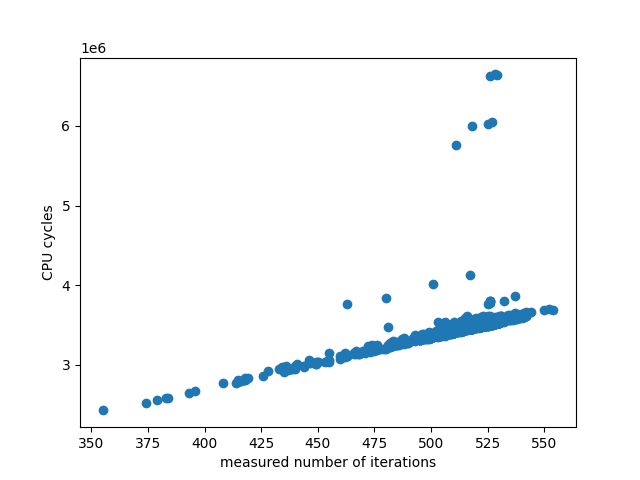}
}\hfil
\subfloat[width=0.47\textwidth][Measured number of iterations vs.\ CPU cycles: noisy trace\label{fig:noisy-trace}]{
\includegraphics[width=0.47\textwidth]{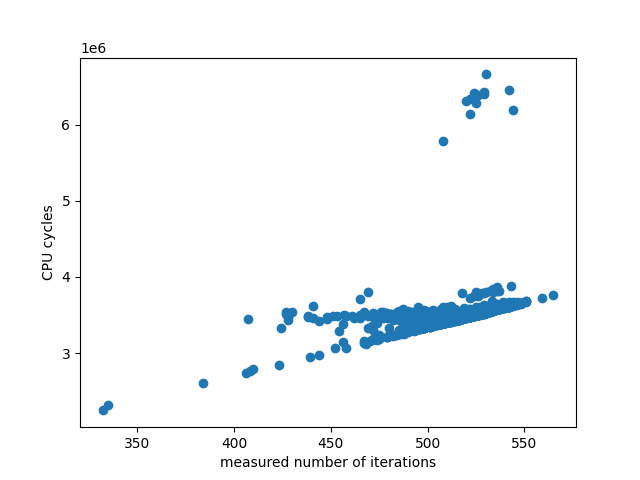}
}\hfil
\subfloat[width=0.47\textwidth][Noise level vs.\ distance to fingerprint\label{fig:noise-vs-distance}]{
\includegraphics[width=0.47\textwidth]{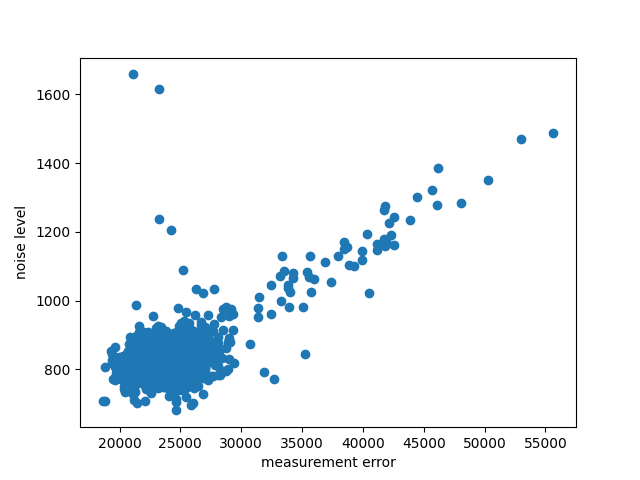}
}\hfil
\subfloat[width=0.47\textwidth][Measurement error after removing noisy traces\label{fig:good-distribution}]{
\includegraphics[width=0.47\textwidth]{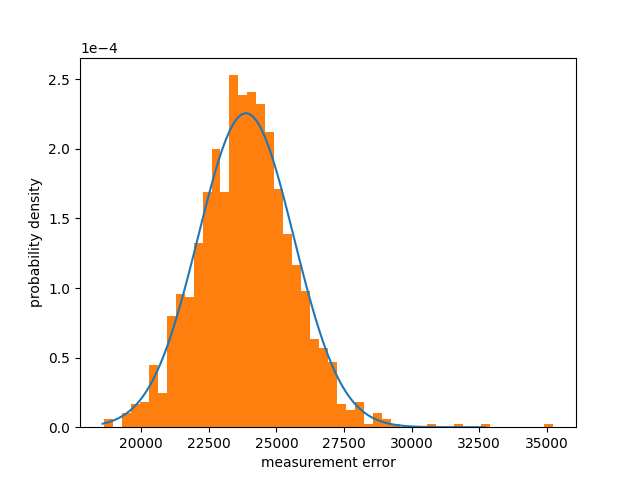}
}

\caption{Filtering out noisy traces.\label{fig:filtering-noisy-traces}}
\end{figure}

The problem is that Flushing and Reloading is two orders of magnitude
slower than executing the target loop.  If the victim performs more than
one iteration per each attacker iteration, the attacker misses accesses.
In our environment, the naive approach only captures a small fraction
of the victim's iterations.  Fortunately, we observe that the fraction
of the victim's iterations captured by the attacker is consistently
around 1.1\%.  Therefore, to estimate the actual number of iterations,
the attacker can simply multiply the measured number by $100/1.1$.

\paragraphbe{Challenge: some traces are very noisy.}
Figure~\ref{fig:noisy-distribution} shows the distribution of measurement
error over 1566 2700-word traces from the reddit-sports dataset.
The distribution has a long tail due to several outliers where the error
is very high.  Figure~\ref{fig:missing-reads} indicates that outliers
are associated with periods when the Flush+Reload loop was slower or
produced more false negatives (an address access by the victim was
not indicated by a lower load time).  This can be due to activity by
concurrent processes sharing the attacker's core, load on the cache bus,
or other low-level interactions.

\label{sec:measurement}

\paragraphbe{Filtering out noisy traces.}
We observe that in a normal state, the execution time of each
iteration of the target loop is usually close to a constant.
Figure~\ref{fig:good-trace} shows the relationship between the measured
number of iterations and time (in CPU cycles).  There are some outliers
(likely caused by CPU interrupts), but the relationship is almost linear.
If a trace is noisy, however, the correlation is weaker\textemdash see
Figure \ref{fig:noisy-trace}.

We measure the ``noise level'' of a trace as the mean squared distance
of its (iterations, time) series relative to the expected line.
Figure~\ref{fig:noise-vs-distance} shows the relationship between the
noise level of a trace and the distance to its corresponding fingerprint.
For our experiments, we removed the 6\% of the traces with the highest
noise levels.  Figure~\ref{fig:good-distribution} shows the histogram of
measurement error after removing these traces.  This histogram fits a
normal distribution model, with symmetry and exponential decay, except
for a few outliers where the measurement error is several standard
deviations away from the mean, indicating that the fit is imperfect.
Even so, the error is always far below $d(N)$, so these outlier traces
would not cause the matching algorithm to produce a false positive.
For higher values of $N$, we expect to also avoid a drop in recall
because the uniqueness radius increases faster than the measurement error
(Figure~\ref{fig:median}).

\section{Data for the case studies}
\label{sec:silkroaddata}

Table~\ref{tbl:silk-road-users} shows variability and measurement error
of the nucleus size series corresponding to the posts of Silk Road
forum users.

\begin{table}[H]
\centering

\begin{filecontents*}{figures/plot_data/silkroad_users.csv}
user,variability,measurement error
cirrus,1546.78,25635.48
indica9,1763.87,22261.53
stealth,1455.91,24129.27
cirrus(SR2),1753.64,20955.76
jediknight,1465.62,22010.80
digitalalch,1586.34,25537.31
synergy,1699.32,22907.05
ssbd,1701.14,23879.66
captainwhitebeard,2004.27,20950.86
colorblack,1502.06,20909.17
nomad bloodbath,2032.92,23724.85
envious,1504.53,22530.60
modziw,1737.35,21170.16
v,1586.50,21646.63
sarge,1522.02,20902.54
warweed,1667.98,22507.60
samesamebutdifferent,1521.18,21508.18
scout,1521.50,22985.98
\end{filecontents*}
\csvautotabular{figures/plot_data/silkroad_users.csv}
\vspace{1em}

\caption{Variability and measurement error for Silk Road Forum users.
($N$ is 2700, $U(N)-d(N)$ is 31860).}
\label{tbl:silk-road-users}
\end{table}

\label{sec:ubuntudata}

\begin{figure}[H]
\centering
\includegraphics[width=0.7\textwidth]{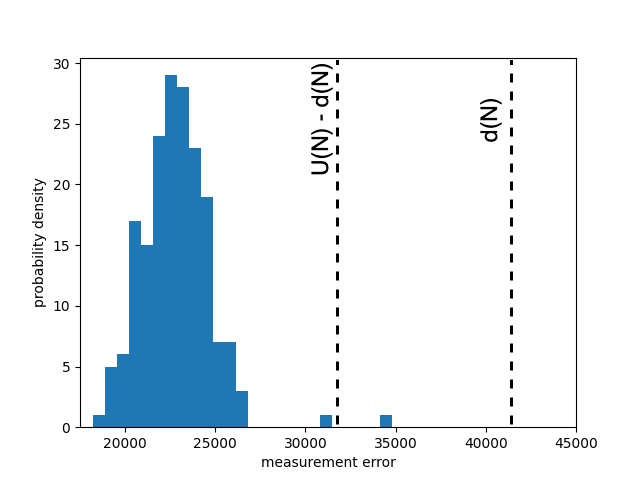}
\caption{Measurement error for Ubuntu Chat users.  ($N$ is 2700,
$U(N)-d(N)$ is 31860, $d(N)$ is 41551).\label{fig:ubuntu}}
\end{figure}

Figure~\ref{fig:ubuntu} shows the distribution of the measurement error
of the nucleus size series corresponding to the posts of the 200 most
active Ubuntu Chat users.  All posts are sufficiently variable.

\end{document}